# Nanomechanical detection of nuclear magnetic resonance using a silicon nanowire oscillator

John M. Nichol[1], Eric R. Hemesath[2], Lincoln J. Lauhon[2], and Raffi Budakian[1]*

[1]Department of Physics, University of Illinois at Urbana-Champaign, Urbana, IL 61801

[2]Department of Materials Science and Engineering, Northwestern University, Evanston, IL 60208

*email: budakian@illinois.edu

**Abstract**

We report the use of a silicon nanowire mechanical oscillator as a low-temperature nuclear magnetic resonance force sensor to detect the statistical polarization of $^1$H spins in polystyrene. Under operating conditions, the nanowire experienced negligible surface-induced dissipation and exhibited a nearly thermally-limited force noise of 1.9 aN$^2$/Hz in the measurement quadrature. In order to couple the $^1$H spins to the nanowire oscillator, we have developed a new magnetic resonance force detection protocol which utilizes a nanoscale current-carrying wire to produce large time-dependent magnetic field gradients as well as the rf magnetic field.

Magnetic resonance force microscopy (MRFM) was proposed as a means of magnetic resonance imaging with the eventual goal of achieving the sensitivity to image individual molecules with atomic spatial resolution[1]. MRFM detects the displacement of an ultrasensitive cantilever in response to the weak force generated by nuclear or electron spins in the presence of a magnetic field gradient. Most recently, MRFM imaging with spatial resolution below 10 nm and sensitivity to fewer than 100 net nuclear spins was demonstrated[2], far exceeding the capabilities of inductively detected magnetic resonance. Extending the resolution and sensitivity to the single proton level will likely require better force sensors, higher magnetic field gradients, or both. "Bottom-up" oscillators such as nanowires[3,4], nanotubes[5,6], and graphene[7] oscillators have been proposed as next-generation force and mass sensors because of their relative ease of fabrication and potential for ultralow mechanical dissipation. In particular, silicon nanowires (SiNWs) have been shown to possess room-temperature force sensitivities[3] far below those of microscale cantilevers and thus offer a promising route to push MRFM closer to achieving the goal of molecular imaging.

The SiNW used in this work was grown epitaxially on a Si[111] substrate using a controlled-diameter vapor-liquid-solid approach with silane as a precursor at 600 °C[8]. The SiNW is approximately 15-μm long with a tip diameter of 35 nm and a base diameter of 200 nm. Figure 1a shows several SiNWs representative of the type used here. The SiNW tip was coated with a thin layer of polystyrene containing $^1$H spins (Fig. 1b). The fundamental flexural mode has a spring constant k = 650 μN/m, a resonance frequency $\omega_0/2\pi$ = 786 kHz, and a quality factor Q = $2.5 \times 10^4$ at 8K.

To detect the displacement of the SiNW, a free-space interferometer coupled to a polarization-maintaining[3] optical fiber was used (Figs. 1c and 1d). The wavelength λ = 2 μm was chosen to minimize optical absorption by the SiNW. The SiNW substrate was mounted on a 3-axis piezoelectric positioner for coarse alignment with respect to the fiber and lens, which are also mounted on a 2-axis piezoelectric bimorph scanner for fine optical alignment (Fig. 1d). The current-carrying wire was mounted on a 3-axis piezoelectric positioner and scanner for coarse and fine positioning with respect to the SiNW. The entire assembly was cooled to 4.2 K in a high vacuum chamber (<$10^{-6}$ mbar). The temperature of the SiNW was approximately 8 K with an incident optical power of less than 1 μW. See Supplementary Information for further details.

Because MRFM requires large magnetic field gradients, the distance between the sample and the gradient source must be very small, typically less than 100 nm. At such small tip-surface separations, the cantilever performance usually degrades considerably due to noncontact friction which arises from surface-induced forces fluctuating at $\omega_0$[9,10]. Noncontact friction presents a serious obstacle to improved sensitivity in MRFM. Remarkably, SiNWs of the type we study maintain their ultralow native dissipation to within 15 nm from a surface (Fig. 2). While the origins of noncontact friction are not in general well understood, the high natural frequency and small cross-sectional area of a SiNW oscillator may help to minimize surface fluctuations.

To take full advantage of the ultralow thermal force noise exhibited by the SiNW, we developed a new MRFM protocol called MAGGIC (Modulated Alternating Gradients Generated wIth Currents). Using this scheme we measure the x-component of the force on the SiNW from the longitudinal component of the spins: $F_x = \mu_z dB_z/dx$, where $dB_z/dx$ is the lateral magnetic field gradient, and $\mu_z$ is the z-component of the spin magnetic moment. The MAGGIC protocol relies on time-dependent currents passing through a small wire[11], which we refer to as the rf wire, to generate both the rf field $B_1$ and the gradient, which has a time-dependence of the form $dB_z(t)/dx = A(t)\cos(\omega_0 t)$ (Fig. 3). To minimize spurious excitation of the SiNW caused by the electric fields produced from the rf wire, the modulation envelope $A(t)$ periodically reverses sign with a period $T_{AM}$ and ensures there is no Fourier component of the voltage across the rf wire at $\omega_0$. (The presence of uncompensated charge on the SiNW causes electric fields oscillating at or near $\omega_0$ to strongly drive the SiNW.) During the time $T_{off}$ when the gradient amplitude is zero, the spins on resonance are inverted adiabatically. Because the spins are reversed synchronously with the envelope $A(t)$, the force generated by the spins does not change sign and resonantly drives the SiNW at $\omega_0$. The signal power at $\omega_0$ is maximized as the gradient duty cycle $D = 1-2T_{off}/T_{AM}$ approaches unity. This requires that $T_{off}$ be as short as possible.

A distinguishing feature of the MAGGIC protocol is the use of electric currents to generate strong, pulsed magnetic field gradients. This capability not only enables nuclear spin MRFM using rf oscillators, but also permits the use of well-established magnetic resonance imaging methods, such as Fourier encoding[12, 13], for efficient data collection. In fact, MRFM using Fourier encoding has previously been proposed[14] and demonstrated[15] with micrometer spatial resolution. The application of these techniques to nanometer scale imaging will be the subject of future work.

To generate the strong time-dependent local fields and gradients for the MAGGIC protocol, we used a focused ion-beam to cut a 375-nm wide and 500-nm long constriction in a larger 500-nm thick Au wire patterned on a Si substrate (Figs. 1e and 1f). The tip of the SiNW was positioned 80 nm above the center of the constriction to maximize both the magnitudes of $B_1$ and $dB_z/dx$. Aproximately 67 mA of current, corresponding to a peak current density of $3.6 \times 10^7$ A/cm$^2$ through the constriction, generated both $B_1$ and the gradient oscillation. A superconducting solenoid provided the static field $B_0 = 0.183$ T along the z-direction. The SiNW was electrostatically damped using a gate electrode (Figs. 1c and 1e) to $Q = 1.3 \times 10^4$ to increase the bandwidth of the oscillator[16]. The spin signal was measured by demodulating the displacement signal using a software-based lock-in amplifier referenced to the gradient oscillation.

Shown in Fig. 4a is the single-sided force spectrum from the in-phase lock-in channel. The MRFM signal from the statistically polarized $^1$H spins in the polystyrene appears as a peak at 0 Hz. The signal power and relaxation time were extracted by fitting the spectrum to a Lorentzian[17]. For typical experimental parameters, the signal power is approximately 6 aN$^2$, and the spin relaxation time is $\tau_m \approx 1$s—a factor of two larger than what has previously been reported in polystyrene[18]. The double-sided force noise in the detection quadrature is $1.9 \pm 0.6$ aN$^2$/Hz. This is slightly above the thermal noise power, which is $1.2 \pm 0.3$ aN$^2$/Hz. The observed force noise is significantly lower than what is measured in microscale cantilevers operating at 300 mK[2]. The uncertainty estimates are based on the uncertainties in the measurements of the quality factor, temperature, and spring constant of the SiNW. We explicitly verified that neither the SiNW thermal amplitude nor its dissipation change upon approach to the operating distance. Therefore, we conclude that the excess force noise of $0.7 \pm 0.4$ aN$^2$/Hz originates from spurious excitation of the SiNW by the rf wire.

To measure the magnitude of $B_1$, we applied nutation pulses (Supplementary Fig. S1) to observe Rabi oscillations (Fig. 4b). From the period of the Rabi oscillations, we find $B_1 = 8.8$ mT. As expected, $\tau_m$ increases with $B_1$ (Fig. 4a upper inset). As a further check, we simulated the magnitude of the signal and the rf field $B_1$ based on the geometry of the rf wire and polystyrene coating; the calculated signal power is 11 aN$^2$ and $B_1 = 9.8$ mT. The agreement with experiment is reasonable given the absence of free parameters in the calculation. The discrepancy may be due to imperfect adiabatic inversions or improper positioning of the SiNW tip over the constriction. Were the SiNW tip actually 200 nm away in the x-direction from the center of the constriction, for example, the calculated signal power is 7 aN$^2$ and $B_1$ = 8.9 mT. Although we have not directly measured the magnetic field gradient, our calculations indicate that $dB_z/dx = 1.2 \times 10^5$ T/m at a distance of 80 nm. Since the gradient falls off rapidly away from the constriction, the polystyrene closest to the SiNW tip contributes most of the signal power (Fig. 4a lower inset). In the future, stronger field gradients in excess of $10^6$ T/m should be possible with smaller constrictions supporting current densities in the $10^9$ A/cm$^2$ range; such large current densities have been reported in nanoscale metal constrictions whose size is small compared with the electron mean-free-path[19].

In summary, we have demonstrated a new route to ultrasensitive MRFM detection using SiNW oscillators and the MAGGIC spin detection protocol. The use of bottom-up NEMS oscillators as force detectors opens the door for greatly improved force sensitivity. Furthermore, the ability to generate large time-dependent field gradients may enable efficient methods for nanoscale magnetic resonance imaging. Together, these new tools promise to advance MRFM closer toward the goal of molecular imaging.

**Acknowledgments**

The authors thank Tyler Naibert for fabrication assistance, and John Mamin and Dan Rugar for helpful discussions. This work was supported by the Department of Physics at the University of Illinois and the Frederick Seitz Materials Research Laboratory. Work at Northwestern University was supported by the National Science Foundation Grant No. DMI- 0507053 through the NIRT program.

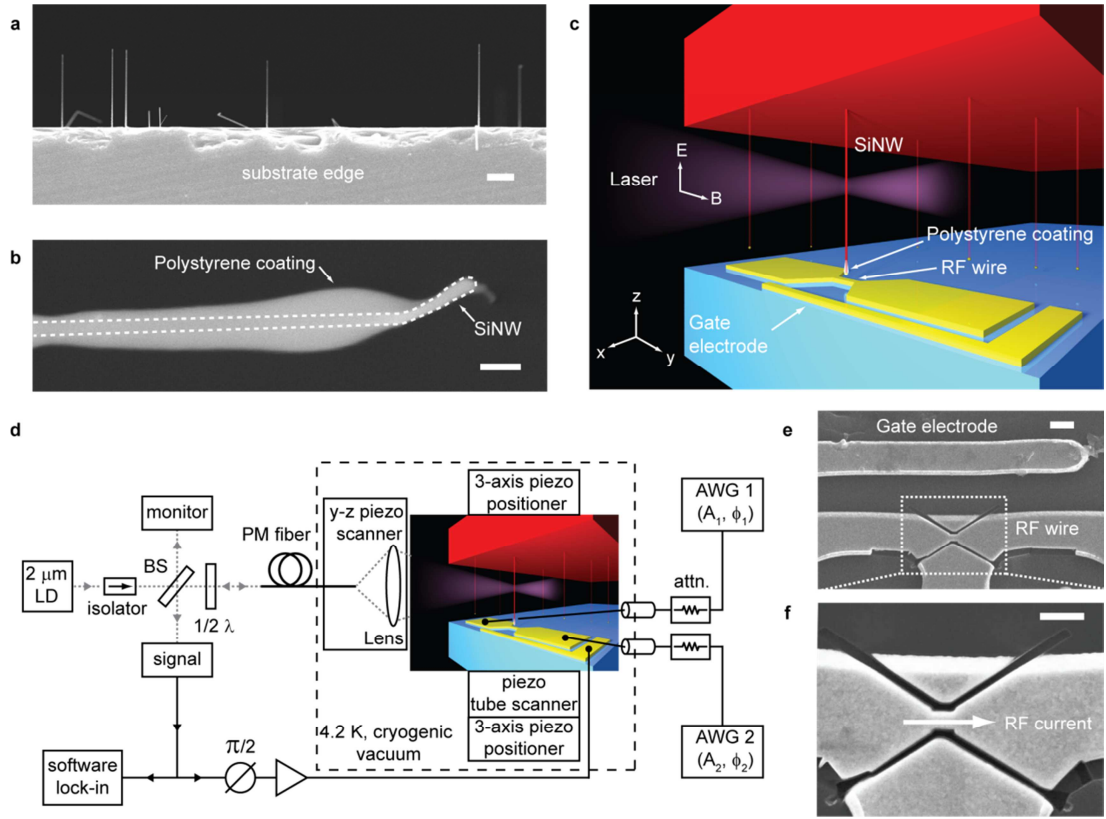

**Figure 1. SiNW and apparatus. a,** Scanning electron micrograph of a SiNW substrate showing several SiNWs representative of the type used in this study. Scale bar is 5 μm. **b,** The tip of the SiNW used in this study with the polystyrene coating. The dashed lines indicate the outer diameter of the SiNW. Scale bar is 100 nm. **c,** Schematic of the experimental setup. Prior to the experiment, a single SiNW on the substrate is selected and coated with polystyrene. The SiNW tip is brought near the constriction in the rf wire. Focused, polarized laser light is used to detect the displacement of the specific SiNW with the polystyrene. **d,** Experimental apparatus. A free-space interferometer coupled to an optical fiber is used to detect the displacement of the SiNW. Light from a 2 μm laser diode (LD) passes through an optical isolator and a beam splitter (BS), couples into a polarization-maintaining (PM) optical fiber, and is focused with a lens onto the SiNW. Two arbitrary waveform generators (AWG1 and AWG2) with independently adjustable amplitudes and phases ($A_1$, $\phi_1$, $A_2$, $\phi_2$) were used to differentially drive the rf wire. **e,** Scanning electron micrograph of the rf wire and gate electrode. Scale bar is 2 μm. **f,** Close up image of the constriction showing cuts made using the focused ion beam. Scale bar is 1 μm.

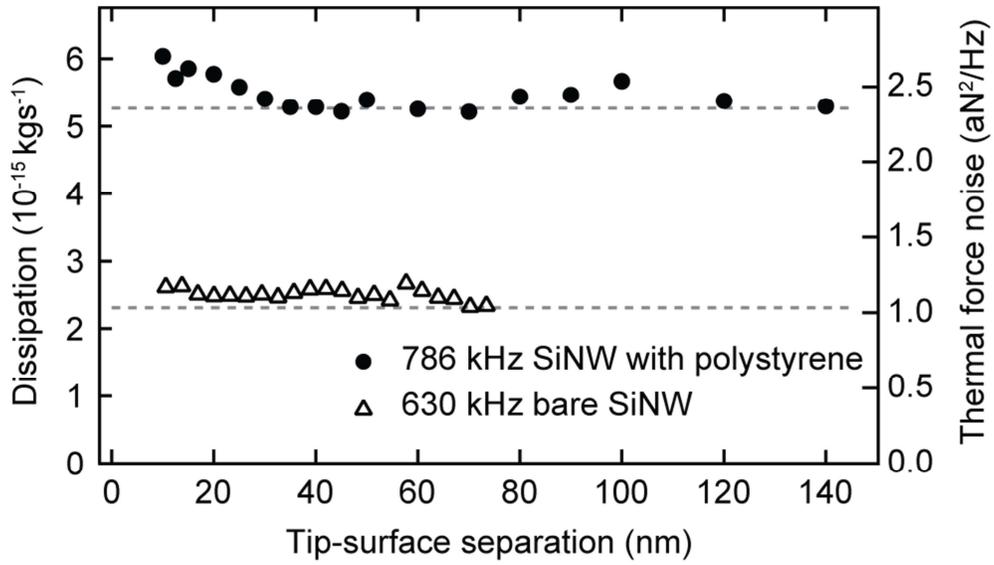

**Figure 2. Surface dissipation.** Total dissipation $\Gamma = k/(\omega_0 Q)$ and thermal force noise $S_F = 4k_B T\Gamma$, where $k_B$ is Boltzmann's constant, of two different SiNWs. In each case, the surface was polycrystalline gold, and the SiNW temperature was T = 8 K. Dashed lines indicate native dissipation as measured far away from a surface. Quality factors were measured by autocorrelation of the oscillator energy. Data are not shown for separations closer than 15 nm as the SiNWs seem to bend slightly upon close approach, making an exact calibration of the distance difficult.

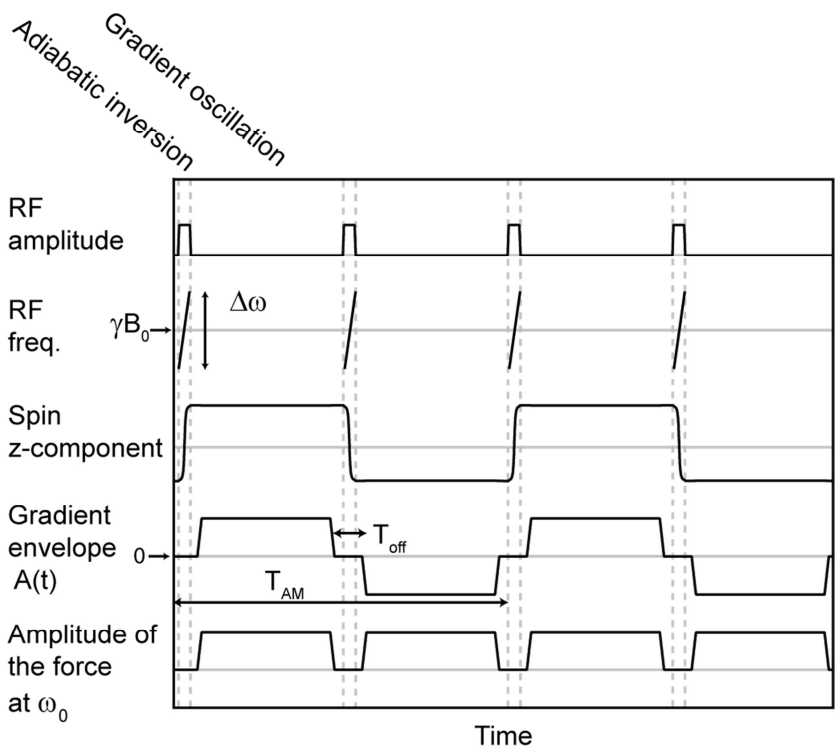

**Figure 3. MAGGIC timing diagram.** $^1$H spins are adiabatically inverted by applying a linear frequency sweep of width $\Delta\omega/2\pi$ = 1.5-3 MHz through resonance $\gamma B_0/2\pi$ = 7.8 MHz at a rate of 20-60 kHz/µs with the rf wire. An AC current at $\omega_0$ generates an oscillating gradient. The gradient oscillation turns off for a duration $T_{off}$, the spins are inverted, and the gradient oscillation turns on again with the opposite sign. The gradient amplitude modulation frequency was typically $f_{AM} = 1/T_{AM}$ = 300-800 Hz, and the duty cycle was typically D ≈ 0.8.

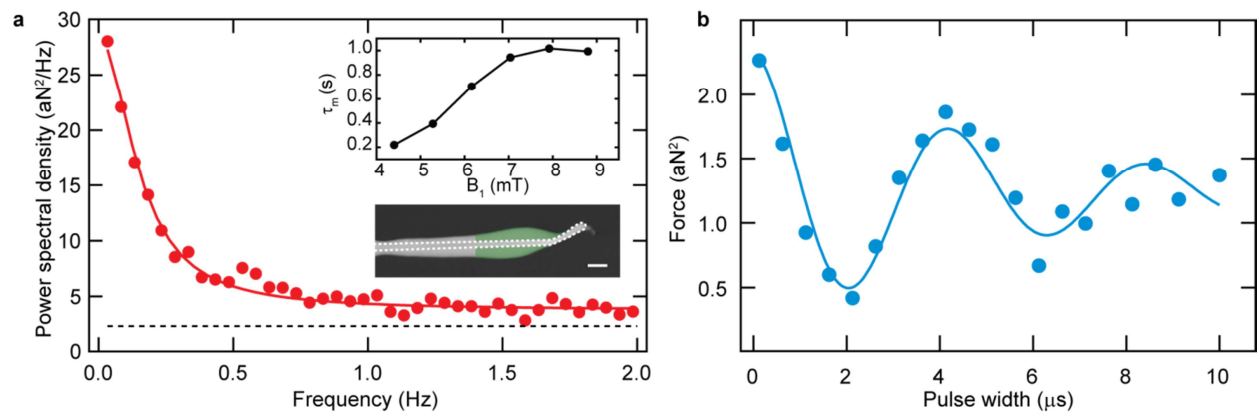

**Figure 4. MRFM data. a,** Spectrum from the software lock-in showing the statistically polarized spin signal and fit to a Lorentzian. The dashed line is the calculated thermal force noise. The shot noise from the interferometer contributes approximately 1 aN$^2$/Hz to the single-sided force noise. This contribution has been subtracted to display only the force noise of the oscillator. Upper inset: Dependence of the spin relaxation time on the rf amplitude. Lower inset: The shaded region of the polystyrene contributes 90% of the observed signal power. Scale bar is 100 nm. **b,** Rabi oscillations and fit to a decaying cosine. See Supplementary Information for more details on the nutation pulse sequence.

# Supplementary Information

# Nanomechanical detection of nuclear magnetic resonance using a silicon nanowire oscillator

John M. Nichol[1], Eric R. Hemesath[2], Lincoln J. Lauhon[2], and Raffi Budakian[1]*

[1]Department of Physics, University of Illinois at Urbana-Champaign, Urbana, IL 61801

[2]Department of Materials Science and Engineering, Northwestern University, Evanston, IL 60208

*email: budakian@illinois.edu

## APPARATUS

A small 4:1 lens (Lightpath 370631) was used to focus the light exiting the optical fiber to a 2.5-μm diameter spot on the SiNW. The working distance of the lens is approximately 300 μm which permits us to conveniently address SiNWs within 20 μm of the substrate edge. Each SiNW substrate contains a random distribution of SiNWs over its surface and is likely to contain several SiNWs suitable for MRFM. Prior to SiNW growth, rulings were etched in the substrate to allow identification and repeated location of individual SiNWs with the optical interferometer. After growth, the substrate was annealed at 400 °C for two hours in forming gas (5% hydrogen and 95% argon) to clean the surface of the SiNW and increase the quality factor.

In order to create the polystyrene coating on the SiNW, a viscous solution of molecular weight 30,000 polystyrene (Pressure Chemical Co. PS80317) dissolved in diethyl-phthalate (Alfa Aesar A17529) was prepared. A drop of the solution was placed on the tip of a glass micropipette and carefully brought into contact with the SiNW under an optical microscope approximately 20 times to build up the coating.

The rf wire was fabricated by sputtering a 5-nm thick Ti/500-nm thick Au film on a silicon substrate with a 500-nm thick layer of thermal oxide. The contact pads and large wires were defined using argon ion milling with a photoresist etch mask. A focused ion beam was used to cut the constriction in the rf wire. Prior to cutting with the focused ion beam, the device was annealed at 250 °C for 3 hours in dry nitrogen to decrease the resistivity of the film. The resistance of the device was approximately 1 Ω at 4.2 K.

**MAGGIC PROTOCOL**

Two arbitrary waveform generators (National Instruments PXI 5412) with independently adjustable amplitudes and phases were used to differentially drive the rf wire. Both generators had nominally the same amplitudes and opposite phases to ensure a voltage null at the constriction and minimize sideband excitation of the SiNW. (Although the voltage across the rf wire contains no Fourier component at $\omega_0$, it does contain sidebands centered about $\omega_0$.) Fine adjustments were made to the amplitude and phase of each generator to further minimize excitation of the SiNW. The current used in this experiment was limited by the compliance of the generators and not by the rf wire, which operated nearly continuously at a current density of $3.6 \times 10^7$ A/cm$^2$ through the constriction for several weeks.

Faster modulation frequencies $f_{AM} = 1/T_{AM}$ were found to improve the signal-to-noise ratio with the same duty cycle, probably because of the reduced sideband excitation of the SiNW. The available magnitude of $B_1$ limited the maximum modulation frequency since faster adiabatic inversions become necessary to keep the duty cycle constant.

To nutate the spins, a pulse sequence similar to previous work[1] was used. Resonant rf pulses of variable length were inserted at regular intervals in the MAGGIC protocol (Fig. S1). The signal power at 0 Hz in the demodulated spectrum was measured as a function of pulse length.

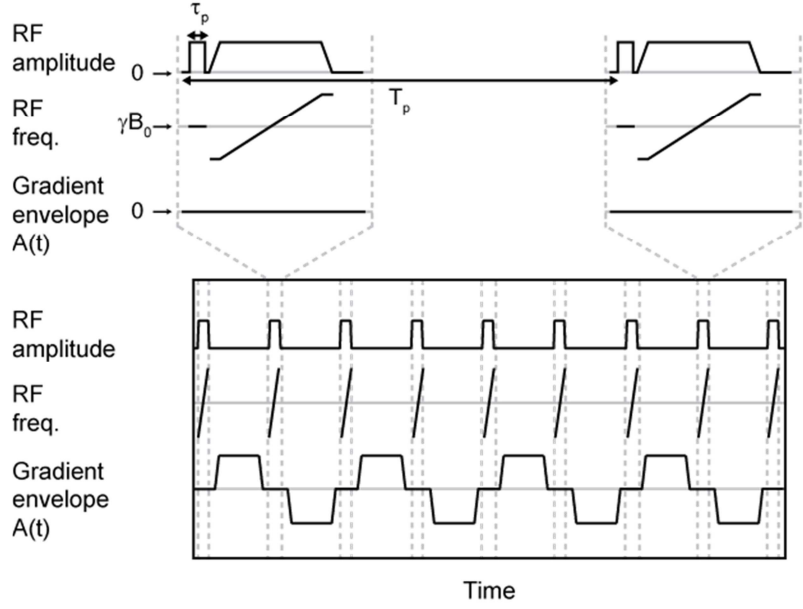

**Figure S1. Nutation pulse sequence.** Rf pulses at $\gamma B_0$ of variable length $\tau_p$ were inserted in the MAGGIC protocol every $T_p = 500$ ms, or every 175 modulation periods at $f_{AM} = 350$ Hz. The period $T_p$ was chosen to be less than the spin lock lifetime $\tau_m$.

**SIMULATION**

To simulate the signal, we modeled the rf wire as an infinitely long wire with a 500 nm × 375 nm rectangular cross section. Finite element analysis using COMSOL Multiphysics (COMSOL, Inc.) of the actual wire geometry including the adjacent metal islands has confirmed that this model reproduces the relevant fields and gradients to within 10% in the region of space occupied by the polystyrene sample. The infinite wire model was used for ease of computation. The shape of the polystyrene coating was extracted from scanning electron micrographs, and the signal power was computed as

$$P = \rho \mu^2 A^2 \cos^2(\theta) \sum_{\mathbf{r} \in V} \Delta V \left( dB_z(\mathbf{r})/dx \right)^2.$$

Here, $\rho = 4.9 \times 10^{28}$ m$^{-3}$ is the $^1$H density in polystyrene, $\mu = 1.4 \times 10^{-26}$ J/T is the proton magnetic moment, $A \approx D/\sqrt{2}$ is the root-mean-square Fourier component of the gradient modulation waveform at $\omega_0$, $\Delta V$ is the volume element of the simulation, and $\theta$ is the tilt angle of the SiNW away from the z-axis (approximately 15 °, as measured with scanning electron microscopy). To calculate the signal power, we summed over all coordinates in the sample volume V. To simulate the Rabi oscillations, the signal power as a function of pulse length $\tau_P$ was simulated as

$$P(\tau_p) = \rho\mu^2 A^2 \cos^2(\theta) \sum_{\mathbf{r} \in V} \Delta V \left(dB_z(\mathbf{r})/dx\right)^2 \left(1 + \cos\left(\gamma B_x(\mathbf{r})\tau_P/2\right)\right)/2.$$

Here, $\gamma = 2\pi \times 42.6$ MHz/T is the proton gyromagnetic ratio. The Rabi frequency was extracted by fitting to an exponentially decaying cosine.